\documentclass{article}

\usepackage[final]{neurips_2023}

\usepackage[utf8]{inputenc} 
\usepackage[T1]{fontenc}    
\usepackage{hyperref}       
\usepackage{url}            
\usepackage{booktabs}       
\usepackage{amsfonts}       
\usepackage{nicefrac}       
\usepackage{microtype}      
\usepackage{xcolor}         
\usepackage{graphicx}
\usepackage{subfig}
\usepackage{multirow}
\usepackage{balance}
\usepackage{nicematrix,enumitem}

\title{COVIDx CXR-4: An Expanded Multi-Institutional Open-Source Benchmark Dataset for Chest X-ray Image-Based Computer-Aided COVID-19 Diagnostics}

\author{%
  Yifan Wu\\
  Department of Systems Design Engineering\\
  University of Waterloo \\
  \texttt{yifan.wu1@uwaterloo.ca} \\
  \And
  Hayden Gunraj\\
  Department of Systems Design Engineering\\
  University of Waterloo \\
  \texttt{hayden.gunraj@uwaterloo.ca} \\
  \And
  Chi-en A. Tai \\
  Department of Systems Design Engineering\\
  University of Waterloo\\
  \texttt{amy.tai@uwaterloo.ca} \\
  \And
  Alexander Wong \\
  Department of Systems Design Engineering\\
  University of Waterloo \\
  \texttt{alexander.wong@uwaterloo.ca} \\
}

\begin{document}

\maketitle

\begin{abstract}
The global ramifications of the COVID-19 pandemic remain significant, exerting persistent pressure on nations even three years after its initial outbreak. Deep learning models have shown promise in improving COVID-19 diagnostics but require diverse and larger-scale datasets to improve performance. In this paper, we introduce COVIDx CXR-4, an expanded multi-institutional open-source benchmark dataset for chest X-ray image-based computer-aided COVID-19 diagnostics. COVIDx CXR-4 expands significantly on the previous COVIDx CXR-3 dataset by increasing the total patient cohort size by >2.66$\times$, resulting in 84,818 images from 45,342 patients across multiple institutions. We provide extensive analysis on the diversity of the patient demographic, imaging metadata, and disease distributions to highlight potential dataset biases. To the best of the authors’ knowledge, COVIDx CXR-4 is the largest and most diverse open-source COVID-19 CXR dataset and is made publicly available as part of an open initiative to advance research to aid clinicians against the COVID-19 disease.
\end{abstract}

\section{Introduction}
COVID-19 is an incredibly contagious disease that was first introduced in late December 2019~\cite{wu2020outbreak}. Even after three years, COVID-19 still has a profound global impact with 9,496 confirmed cases in Canada in the two weeks (August 27 to September 9, 2023)~\cite{Covid-Stats}. Subsequently, there has been immense research focused on effective screening of COVID-19 with many deep learning models proposed for chest X-ray (CXR) imaging~\cite{al2023covid, hussain2021corodet,rahman2021exploring}. The usage of CXR increased as a complement to RT-PCR testing as CXR minimizes the risk of cross-infection \cite{jacobi2020portable} and is routinely used for respiratory concerns in clinical use cases \cite{nair2020british}. However, the diversity and scale of COVID-19 CXR datasets have limited the advancements of model improvements for public usage. In this paper, we introduce COVIDx CXR-4, an expanded multi-institutional open-source benchmark dataset for chest X-ray image-based computer-aided COVID-19 diagnostics. COVIDx CXR-4 expands significantly on the previous COVIDx CXR-3 dataset by increasing the total patient cohort size by >2.66$\times$, resulting in 84,818 images from 45,342 patients across multiple institutions. We provide extensive analysis on the diversity of the patient demographic, imaging metadata, and disease distributions to highlight potential biases in the dataset. To the best of the authors’ knowledge, COVIDx CXR-4 is the largest and most diverse open-source COVID-19 CXR dataset. COVIDx CXR-4 is part of an open initiative to advance research to aid clinicians against the COVID-19 disease and is made publicly available at \url{https://www.kaggle.com/datasets/andyczhao/covidx-cxr2}.

\begin{figure}
    \centering
{\includegraphics[width=\linewidth]{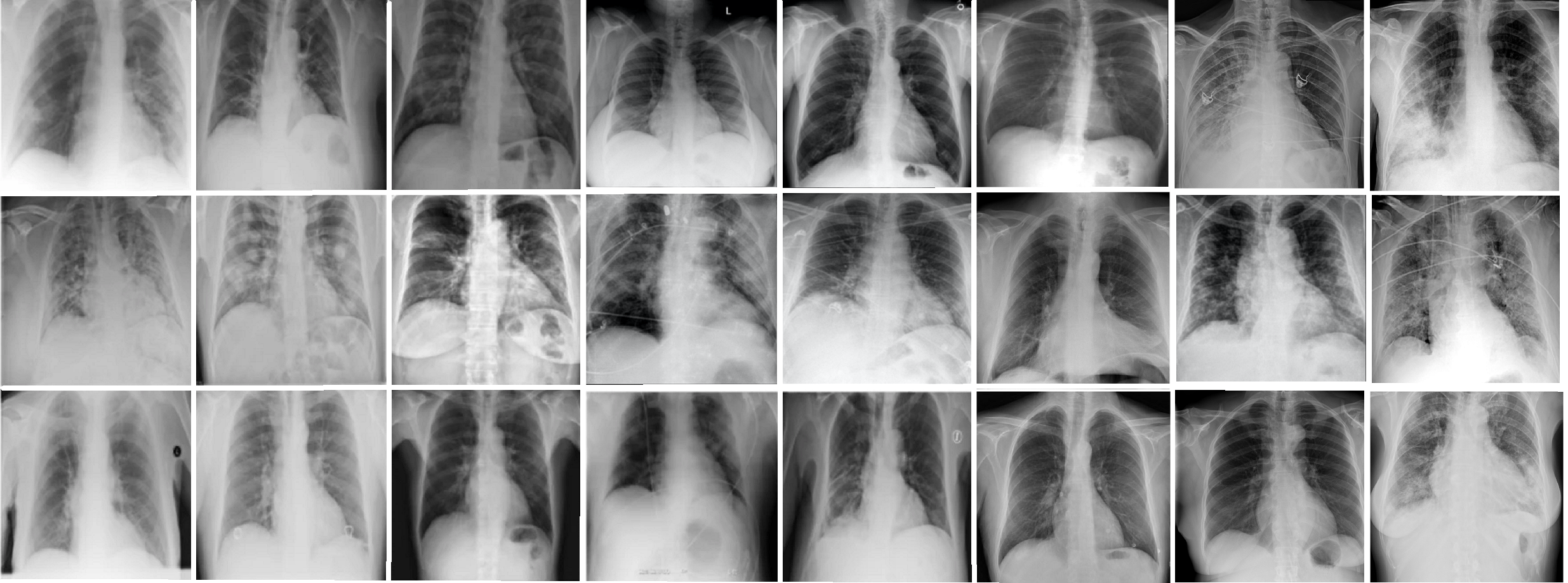}}
    \caption{Example CXR images from patient cohort in COVIDx CXR-4 dataset.}
    \label{fig:demo-dist}
\end{figure}

\section{Methodology}
The COVIDx CXR-4 benchmark dataset expands significantly upon the COVIDx CXR-3 benchmark dataset~\cite{pavlova2022covidx}, including the addition of data that was carefully curated from BIMCV-COVID19+ and BIMCV-COVID19-~\cite{bimcv, w3aw-rv39-21, m4j2-ap59-21}. Images with incomplete meta information were filtered out and both min-max normalization and white/black inversion were applied to the images according to the meta information. Then images in lateral view were filtered out and images with inappropriate characteristics (e.g., wrong clip range, wrong orientation, inverted white and black, containing other body parts, failed X-ray images) were also removed. Hashes were compared for duplicate images, which were also discarded from the dataset.

\begin{figure*}[!ht]
    \centering
    \subfloat[CoV+ Patient 1]{\includegraphics[width=.3\linewidth]{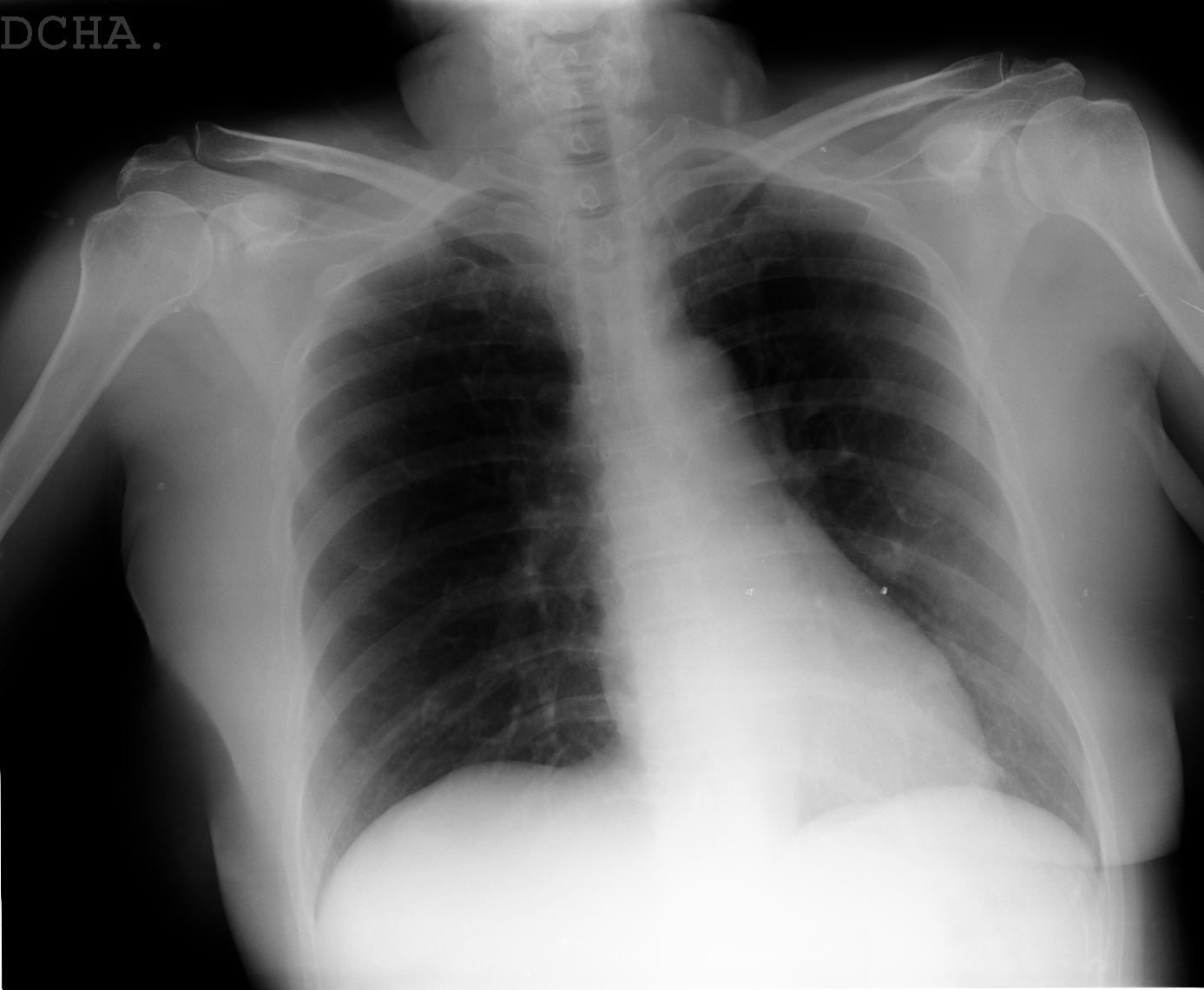}}
    \hfil
    \subfloat[CoV+ Patient 2]{\includegraphics[width=.3\linewidth]{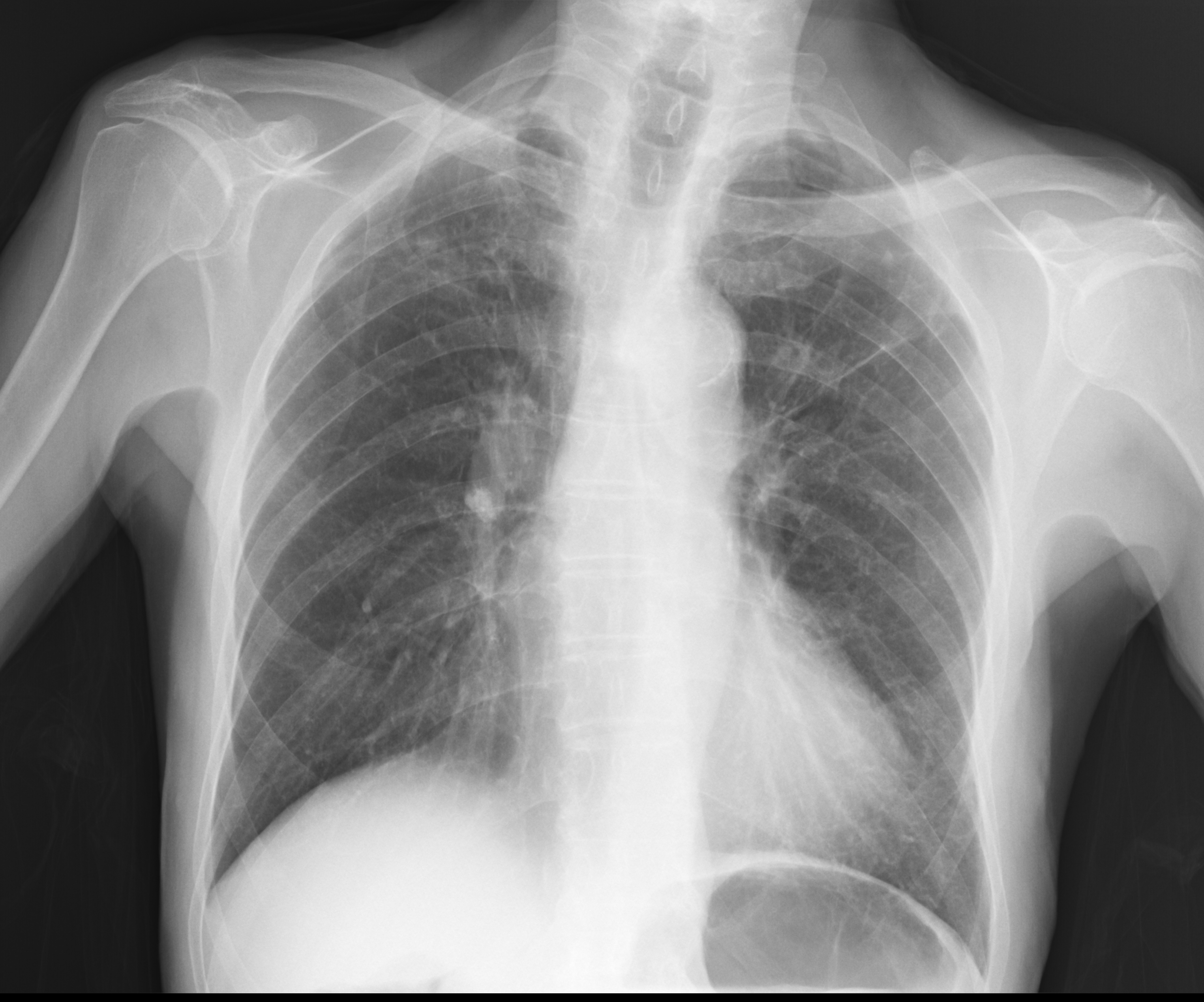}}
    \hfil
    \subfloat[CoV+ Patient 3]{\includegraphics[width=.3\linewidth]{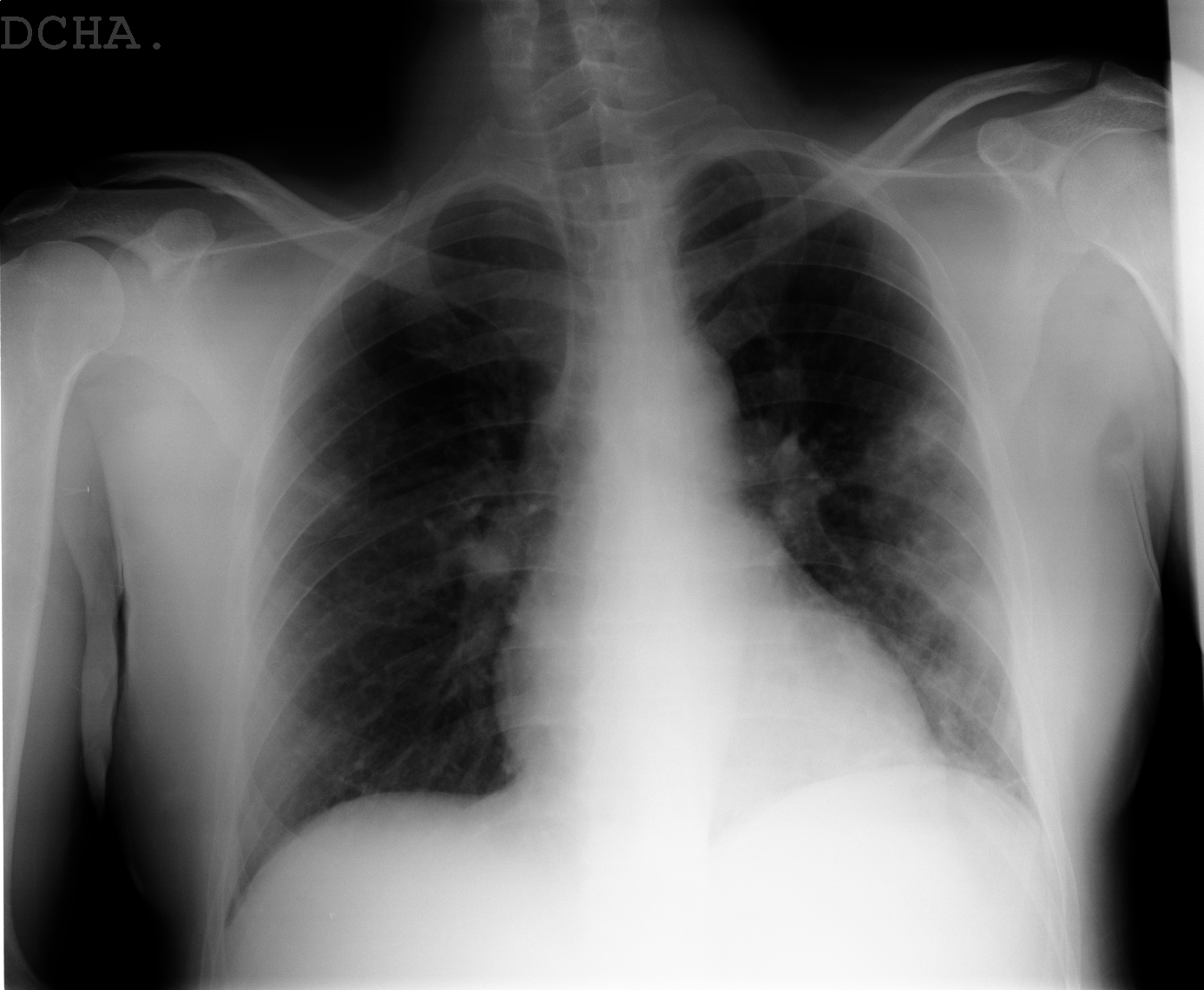}}
    \hfil
    \caption{Example positive case patient images from the COVIDx CXR-4 dataset.}
    \label{fig:example-pos-cases}
\end{figure*}

\begin{figure*}[!ht]
    \subfloat[CoV- Patient 1]{\includegraphics[width=.3\linewidth]{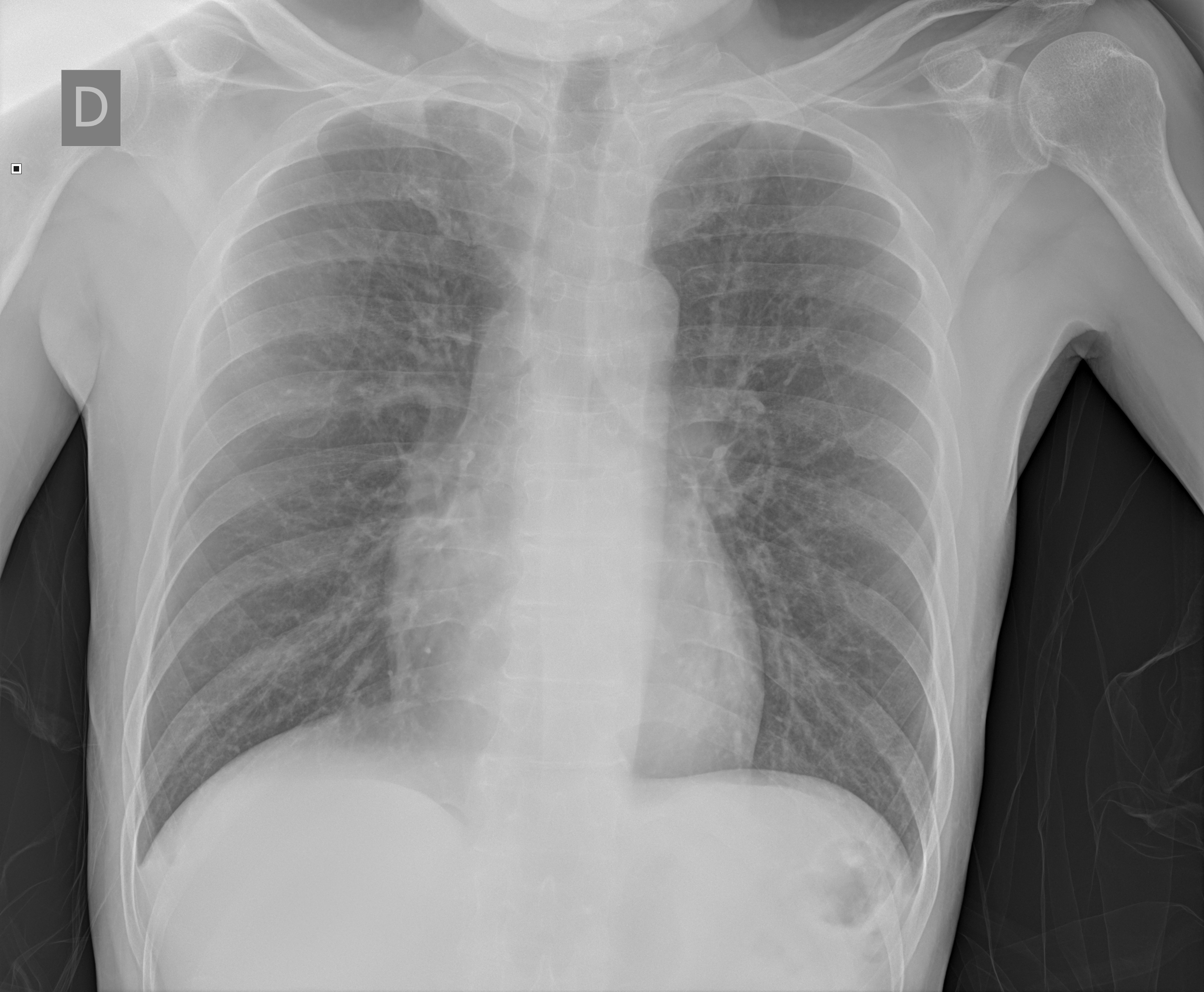}}
    \hfil
    \subfloat[CoV- Patient 2]{\includegraphics[width=.3\linewidth]{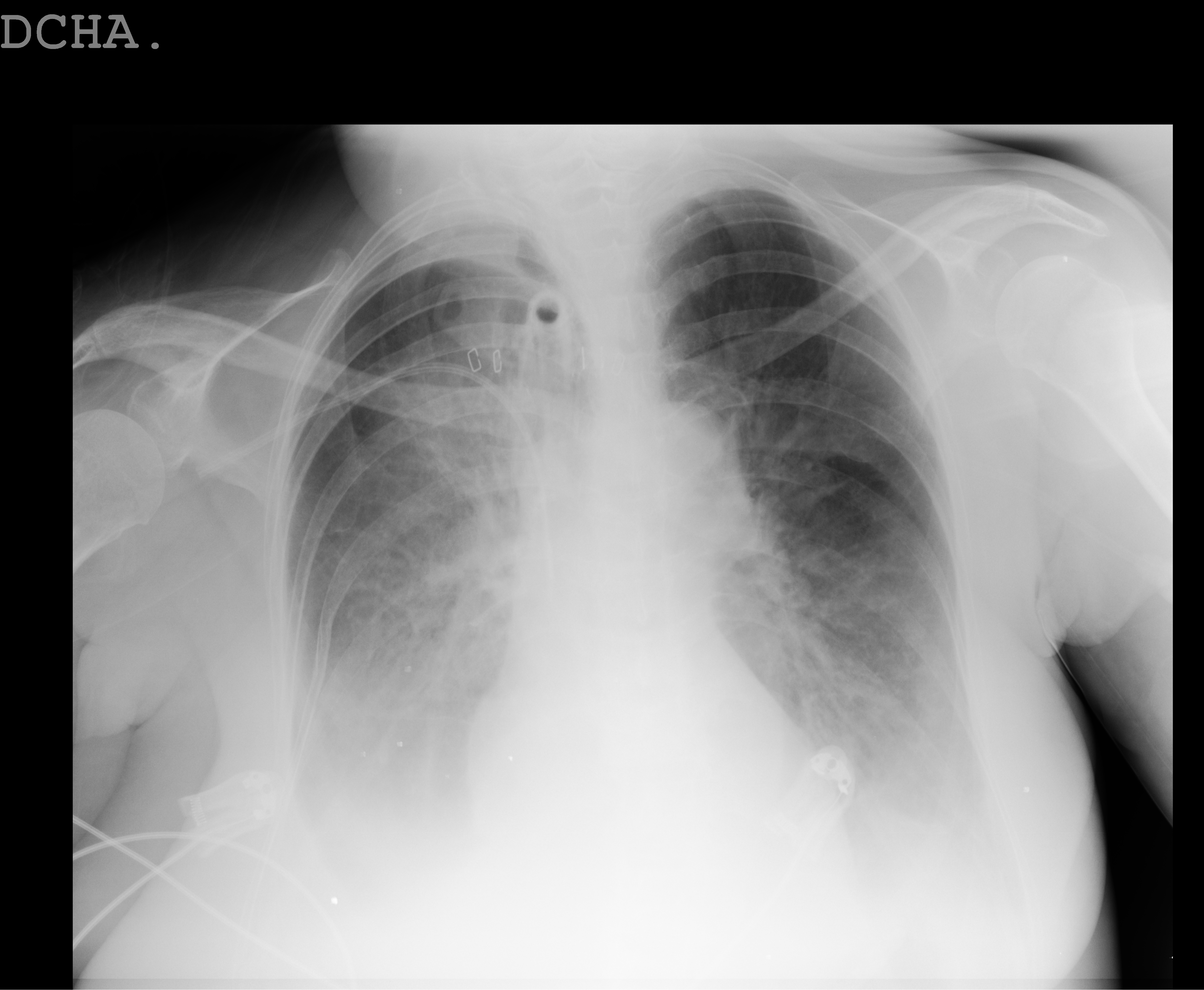}}
    \hfil
    \subfloat[CoV- Patient 3]{\includegraphics[width=.3\linewidth]{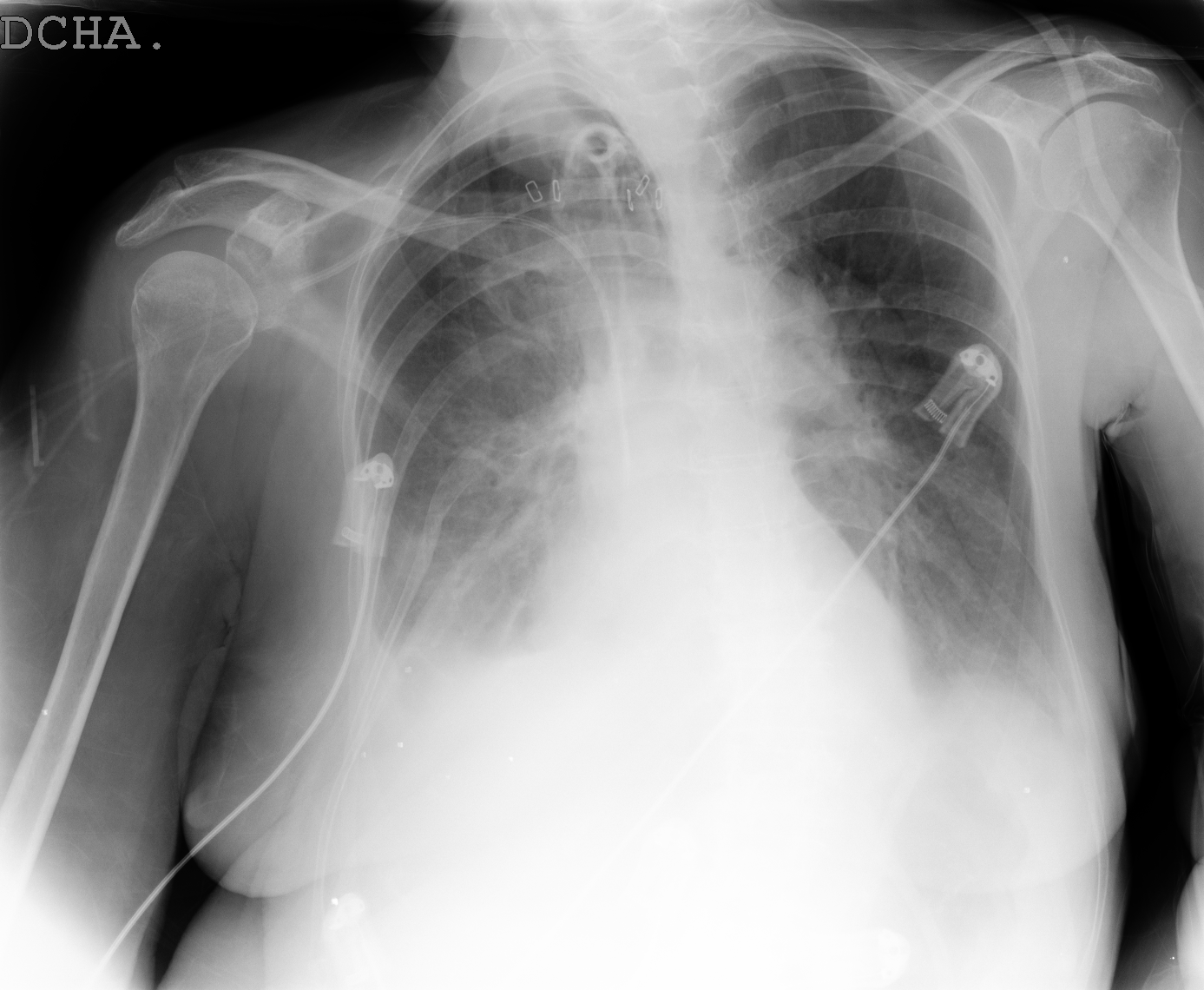}}
    \hfil
    \caption{Example negative case patient images from the COVIDx CXR-4 dataset.}
    \label{fig:example-neg-cases}
\end{figure*}

The training, validation, and test set of the COVIDx CXR-4 dataset contain 80\%, 10\%, and 10\% of all data respectively. The BIMCV data was used to expand the test set from the COVIDx CXR-3 dataset and some of the training data was redistributed into a validation set. Specifically, the COVIDx CXR-4 test set has 10\% of all the data (COVIDx CXR-3 and BIMCV) and is balanced between both positive and negative samples. The remaining BIMCV data and COVIDx CXR-3 training set were combined and split into a training and validation set. The validation set has the same size as the test set (10\% of all the data) and is also balanced between positive and negative samples. On the other hand, the training set is not balanced and consists of the remaining data. The sets were split at the patient level rather than the image level with a 1:8 ratio for validation:training for positive and negative samples respectively for BIMCV and every data source in COVIDx CXR-3. Example images from COVIDx CXR-4 for positive and negative infection cases are shown in Figure~\ref{fig:example-pos-cases} and Figure~\ref{fig:example-neg-cases} respectively.

\section{Results}
Table~\ref{tab:infection-distribution} shows the infection type distribution of CXR images and patient cases in the COVIDx CXR-4 dataset. The final COVIDx CXR-4 benchmark dataset contains 84,818 CXR images from 45,342 patients and contains balanced validation and test sets. However, there is class imbalance in the training dataset as there are roughly five times more positive samples than negative which could result in potentially biased models. Though the system would be evaluated on balanced data (validation and test), we recommend using strategies such as data sampling and balanced loss functions to mitigate the effects of class imbalances when training deep learning models. 

\begin{table}[!ht]
  \caption{Infection type distribution of CXR images and patient cases (in parentheses).}
  \centering
  \NiceMatrixOptions{notes/para}
    \begin{NiceTabular}{l c c c}
        \toprule
        \RowStyle{\bfseries}
        \multirow{2}{*}{Split} & \multicolumn{2}{c}{Infection Type} & \multirow{2}{*}{Total} \\
        & Positive & Negative & \\ \midrule
    Training & 57,199 (24,978) & 10,664 (10,489) & 67,863 (35,457) \\
    Validation & 4,241 (1,010) & 4,232 (4,153) & 8,473 (5,163) \\
    Test & 4,241 (1,589) & 4,241 (3,133) & 8,482 (4,722) \\
    \bottomrule
    \end{NiceTabular}
  \label{tab:infection-distribution}
\end{table}

\begin{table}[!ht]
  \caption{Summary of age demographic in the COVIDx CXR-4 dataset.}
  \centering
  \NiceMatrixOptions{notes/para}
    \begin{NiceTabular}{l r r}
        \toprule
        \RowStyle{\bfseries}
    Age & Number of Patients & Percentage \\ \midrule
    <18 & 1,415 & 3.1\% \\
    [18, 59] & 24,142 & 53.2\% \\
    (59, 74] & 10,232 & 22.6\% \\
    (74, 90] & 7,405 & 16.3\% \\
    >90 & 1,007 & 2.2\% \\
    Unknown & 1,141 & 2.5\% \\
    \bottomrule
    \end{NiceTabular}
  \label{tab:age-demographics}
\end{table}
As seen in Table~\ref{tab:age-demographics}, the patients have a wide age range from less than 18 to greater than 90 years of age. However, over half of the patient cohort (53.2\%) are between 18 to 59 years old which shows a potential bias for this category. Additionally, only 3.1\% of the dataset is below 18 years which signifies significant underrepresentation for this age range in the data.

\begin{figure}[!ht]
    \centering
    \subfloat[Sex (Patient Level)]{\includegraphics[width=.5\linewidth]{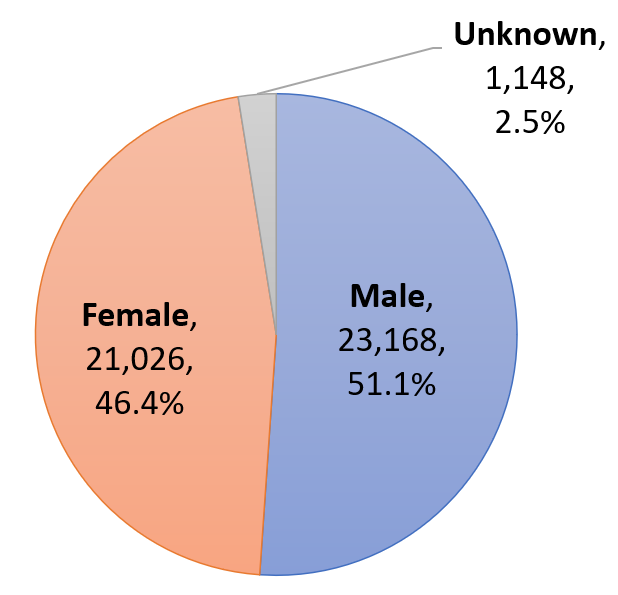}}
    \hfil
    \subfloat[Imaging View (Image Level)]{\includegraphics[width=.45\linewidth]{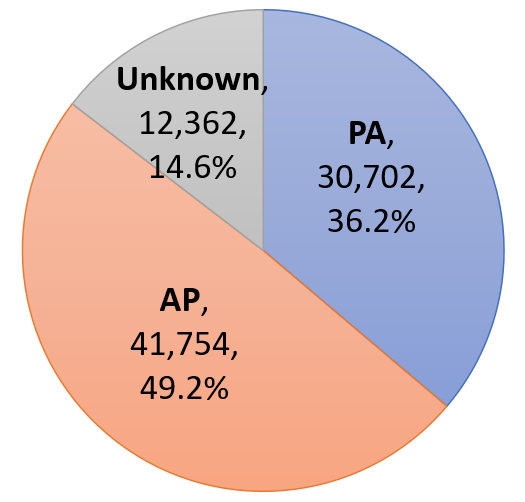}}
    \caption{Sex (patient level) and imaging view (image level) distributions in  COVIDx CXR-4.}
    \label{fig:sex-dist}
\end{figure}

On the other hand, as seen in Figure~\ref{fig:sex-dist} (a), the patient sex category is fairly well balanced with roughly equal proportion of female and male patients (only a 5\% difference in size). Notably, there is some unknown data so it may skew the balance of the sex category but it will likely have minimal impact as unknown only represents 2.5\% of the dataset. The imaging view on an image level presents a different story however as there are more anterior-posterior (AP) images (49.2\%) compared to posterior-anterior (PA) images (36.2\%) as seen in Figure~\ref{fig:sex-dist} (b). Though there is a significant number of images in the unknown category (14.6\%), it is hard to determine if those images would create more or less balance for the imaging view aspect. 

With the data imbalance in the age and imaging view categories, it is important that users of models developed using this data are cognizant of the potential biases that these models may possess. For instance, having less data for patients who are less than 18 years old may result in a lower model accuracy for these patients compared to those in the 18 to 59 age group.

\section{Potential Negative Societal Impact}
Potential negative societal repercussions include the improper use of the data and an excessive reliance on models trained with this dataset. While the primary motive behind releasing this benchmark dataset for public use is to support advancements in research within this domain, there is a risk that others might exploit the data to create algorithms that adjust insurance premiums based on projected medical expenses for individual patients. Conversely, utilizing the data to train models for clinical purposes could also have detrimental effects, especially if there is an undue dependence on the model's results without adequate validation or continuous retraining. Consequently, we strongly emphasize the importance of validating any models trained with this dataset using real-world clinical data and recommend their use under the guidance of experts.

\begin{ack}
The authors thank the Natural Sciences and Engineering Research Council of Canada and the Canada Research Chairs Program.
\end{ack}

{
\small

\bibliography{neurips_2023}
}

\end{document}